\documentclass[twocolumn,amsthm]{autart}    
\usepackage{amsmath,amssymb,amsfonts,amstext}
\usepackage{graphicx}
\usepackage{multirow,tabularx}

\usepackage{todonotes}

\usepackage{bm}
\usepackage{float}
\usepackage{mathtools}

\newtheorem{theorem}{Theorem}
\newtheorem{lemma}{Lemma}

\newtheorem{corollary}{Corollary}

\newtheorem{remark}{Remark}

\newtheorem{proposition}{Proposition}

\newtheorem{problem}{Problem}

\usepackage[round]{natbib}%
\usepackage{algorithm}
\usepackage{algorithmicx} 
\usepackage{algpseudocode} %

\usepackage[position=bottom]{subfig}

\usepackage{hyperref}
\hypersetup{
  colorlinks=true,
    linkcolor= blue,
    allcolors=blue,
    citecolor = blue,
    filecolor=black,      
    urlcolor=blue,
    }

\begin{document}

\begin{frontmatter}

\title{Optimal trajectory planning meets network-level routing:\\Integrated control framework for emerging mobility systems}

\thanks[footnoteinfo]{This research was supported by NSF under Grants CNS-2149520 and CMMI-2219761.}

\author[UD,Cornell]{Heeseung Bang}\ead{hb489@cornell.edu},   
\author[Cornell]{Andreas A. Malikopoulos}\ead{amaliko@cornell.edu}               

\address[UD]{Department of Mechanical Engineering, University of Delaware,\\126 Spencer Lab, 130 Academy St, Newark DE 19716, United States of America}                  
\address[Cornell]{School of Civil and Environmental Engineering, Cornell University,\\ 220 Hollister Hall, 527 College Ave, Ithaca, NY 14853, United States of America}


\begin{abstract}
In this paper, we introduce a hierarchical decision-making framework for emerging mobility systems. Despite numerous studies focusing on optimizing vehicle flow, practical feasibility has often been overlooked. To address this gap, we present a route-recovery method and energy-optimal trajectory planning tailored for connected and automated vehicles (CAVs) to ensure the  realization of optimal flow. Our approach identifies the optimal vehicle flow to minimize total travel time while considering consistent mobility demands in urban settings. We deploy a heuristic route-recovery algorithm that assigns routes to CAVs and departure/arrival time at each road segment. Furthermore, we propose an efficient coordination method that rapidly solves constrained optimization problems by flexibly piecing together unconstrained energy-optimal trajectories. The proposed method has the potential to effectively generate optimal vehicle flow, contributing to the reduction of travel time and energy consumption in urban areas.
\end{abstract}

\end{frontmatter}

\section{Introduction} \label{sec:introduction}

The urban traffic landscape has significantly changed due to rapid urbanization and the continuous growth of the global population.
These rapid changes have created a complex traffic network and broader challenges, including pollution, inequity in mobility, congestion, and environmental impacts.
In response to these challenges, emerging mobility systems, including connected and automated vehicles (CAVs), shared mobility, and autonomous mobility-on-demand (AMoD) systems, have emerged as promising solutions.
These systems not only hold the potential to address current challenges but also promise to enhance safety, reduce user costs, and optimize traffic network efficiency.

Researchers have explored various approaches to leverage CAVs to create safe and efficient driving environments in this context.
For instance, several studies have addressed eco-driving in signalized intersections \citep{ma2021trajectory,han2020energy}, cooperative adaptive cruise control \citep{gong2019cooperative,wang2020cooperative}, efficient brake control \citep{hu2022distributed}, and highway speed harmonization \citep{malikopoulos2018optimal}. Some research efforts have delved into the coordination challenges of CAVs in different traffic scenarios, such as merging on ramps \citep{Rios-Torres2,liu2023safety,davarynejad2011motorway}, traversing roundabouts \citep{martin2021traffic} or signal-free intersections \citep{Malikopoulos2017,xu2021comparison,Malikopoulos2020}, safe navigation in mixed traffic settings \citep{ghiasi2019mixed,li2022simulation}, and platoon formation \citep{han2020energy}.

Beyond the scope of these low-level control approaches, several research efforts have also ventured into higher-level decision-making. Numerous studies have explored routing strategies for AMoD systems \citep{pavone2012robotic,zardini2022analysis,spieser2014toward}. In contrast, others have considered issues related to battery limitations \citep{ding2021integrated} or ride-sharing options in the routing problem \citep{alonso2017predictive,wallar2018vehicle,fielbaum2021demand}.

Despite the considerable body of work addressing decision-making challenges in emerging mobility systems across various levels, a critical gap has yet to be addressed: the lack of integration across these diverse research efforts.
This paper aims to bridge this gap by introducing a multi-level decision-making framework to provide optimal operational strategies at each level.
This framework can be used as a roadmap, fostering a comprehensive approach to address the complexities of emerging mobility systems and connecting disparate research efforts into a unified strategy for more effective and holistic solutions.

\subsection{Related Work} \label{subsec:literature}
CAVs have attracted significant research attention due to their potential to enhance transportation efficiency and alleviate congestion.
Numerous studies have addressed time and energy-efficient coordination challenges in diverse traffic scenarios, including merging roadways \citep{liu2023safety}, lane changes \citep{duan2023cooperative}, roundabouts \citep{alighanbari2020multi}, signal-free intersections \citep{Au2015,Malikopoulos2020,kuchiki2022cooperative,zhang2021trajectory}, and corridors \citep{lee2013,chalaki2020TITS,mahbub2020decentralized}.

Efforts to optimize CAV operations and improve efficiency have resulted in eco-driving control at signalized intersections \citep{sun2020optimal}, safety measures in roundabout crossings and mixed traffic merging scenarios using control barrier functions \citep{abduljabbar2021control,hamdipoor2023safe}, and energy-efficient speed planning for connected and automated electric vehicles \citep{wang2022energy}.
Other studies have explored platooning in mixed traffic scenarios \citep{mahbub2023_automatica} and collision avoidance during CAV platoon merging and splitting \citep{wang2023collision}.
Furthermore, other research efforts have examined CAV intersection dynamics and learning \citep{luo2023real}, as well as mixed traffic scenarios considering social value orientation \citep{toghi2021cooperative,Le2022CDC}.
Meanwhile, several efforts have focused on utilizing learning to incorporate human-driver behavior or address the problem's uncertainty. For example, \citet{Nishanth2023AISmerging} employed approximate information states for human-driven vehicle modeling in mixed-traffic merging scenarios, while \citet{wang2023coordination} explored intersection crossing with a learning perspective.

Other research efforts have focused on routing CAVs with various strategies to mitigate computational challenges.
\citet{chu2017dynamic} presented a routing method using a dynamic lane reversal approach. To consider the influence of other CAVs, the authors estimated travel time proportional to the number of CAVs on the roads. Similarly, \citet{mostafizi2021decentralized} designed a heuristic algorithm for a decentralized routing framework, which estimates the travel speeds inversely proportional to the number of CAVs on the same road. As another way to avoid computational issues, a series of papers addressed the decentralized routing problem using the ant colony optimization technique, which uses a metric representing the traffic condition and updates the information in real-time \citep{bui2019aco,nguyen2021ant,nguyen2021multiple}.
From a microscopic perspective, many studies have explored single-vehicle routing, optimizing routes based on the surrounding environment \citep{de2017bi,houshmand2021combined,boewing2020vehicle,gammelli2021graph}. Some approaches consider sequential order \citep{Bang2022combined}) or person-by-person optimal routing among CAVs \citep{Bang2022rerouting}.

While these studies contribute valuable insights, a gap remains in scaling up for large deployments.
This led to attention to the vehicle flow optimization for AMoD systems on the macroscopic level.
Several studies focused on minimizing travel time while addressing vehicle relocation \citep{houshmand2019penetration,salazar2019congestion}, mixed traffic environments \citep{wollenstein2020congestion,wollenstein2021routing}, and charging constraints \citep{bang2021AEMoD}.
Although these studies employed travel latency functions to capture the influence of other vehicles, practical implementation of this optimal flow remains a challenge.

There also have been other efforts on the efficient management of large-scale AMoD systems using discrete-time models and model predictive control \citep{carron2019scalable}, or on considering ride-sharing option using predictive control strategies \citep{alonso2017predictive,tsao2019model} and learning approaches \citep{gueriau2020shared}.
Additionally, charging scheduling for shared AMoD systems has been addressed \citep{liang2020mobility}.

\begin{figure*}
    \centering
    \includegraphics[width=0.7\linewidth]{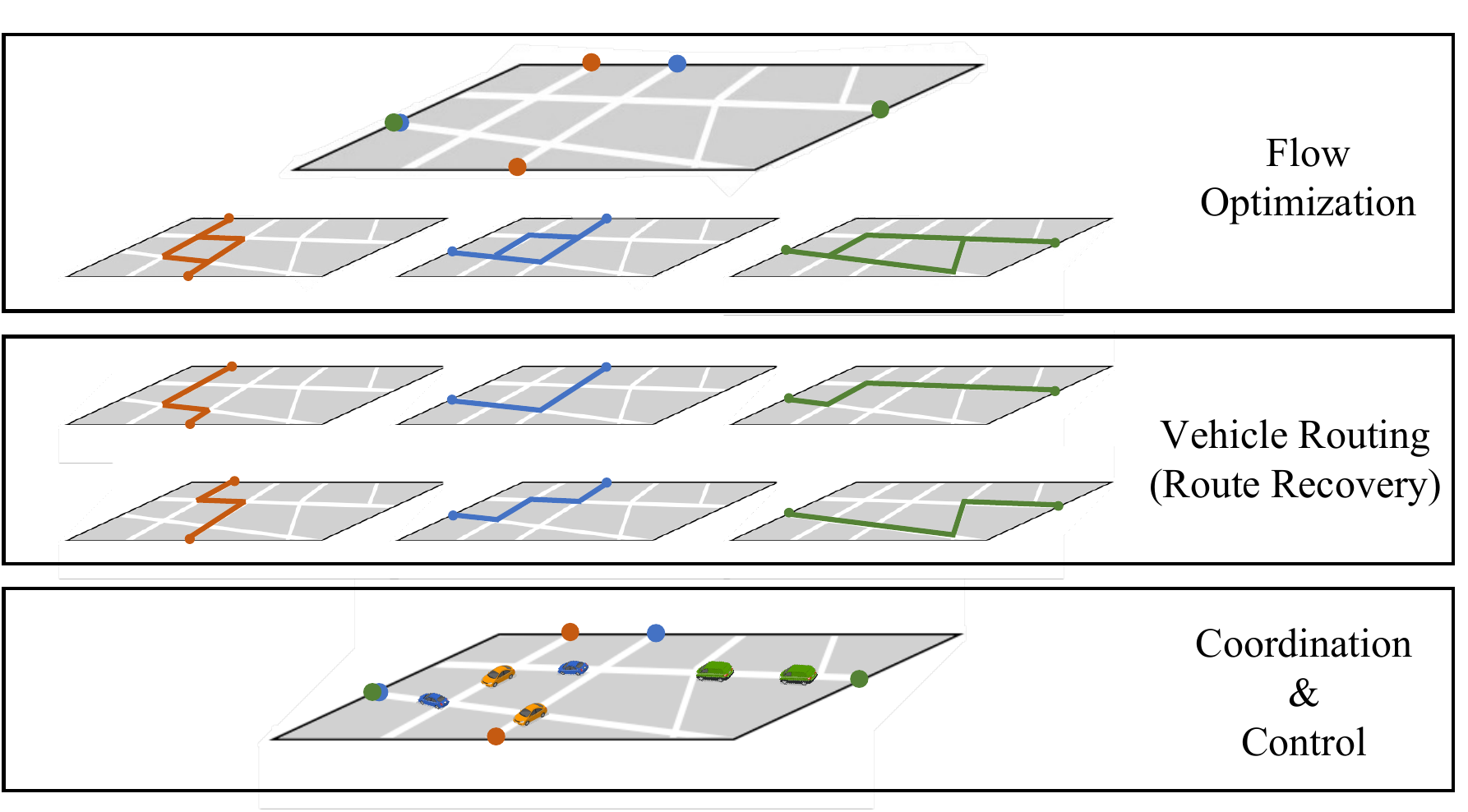}
    \caption{A hierarchical structure for flow-based routing and coordination.}
    \label{fig:structure}
\end{figure*}

\subsection{Contributions} \label{subsec:contribution}

Despite the extensive research in developing optimal control methods for optimizing the efficiency of CAVs and AMoD system management, there is a notable gap in integrating these domains and analyzing the mutual impact of their solutions.
This paper presents a hierarchical decision-making framework that provides optimal operational strategies across multiple levels.
Our research introduces several key contributions summarized as follows:

\textit{1) Introduction of a hierarchical decision-making framework:}
Our work addresses the intricate decision-making requirements of emerging mobility systems by integrating processes at various levels, as illustrated in Fig. \ref{fig:structure}.
Starting with an upper-level flow-optimization problem for AMoD systems, the framework extracts routes from the obtained flow and assigns them to CAVs with appropriate departure and arrival times.
Subsequently, considering energy consumption and safety constraints, CAVs plan their trajectories at traffic scenarios, e.g., intersections, merging roadways, and speed reduction zones.
If any of these trajectories become inapplicable, we resolve the routing problem by incorporating the infeasible feedback from the vehicle coordination  at the low level.

\textit{2) Derivation of optimality conditions for vehicle-level trajectory planning:} 
Our approach establishes the necessary optimality conditions for vehicle-level trajectory planning.
The analysis of these conditions provides a critical foundation for understanding vehicle-level control strategies in flow-based CAV operations. We identify the conditions under which these controls are optimal and offer insights into what actions to take when optimality cannot be achieved.

\textit{3) Development of a heuristic approach for practical implementation:}
We introduce a practical heuristic approach designed to implement optimal traffic flow in the context of CAV operations. Recognizing the limited attention to this critical aspect, our heuristic serves as a preliminary method for practical applications, enabling the realization of optimal traffic flow in real-world scenarios.

We recently conducted a limited-scope analysis of the hierarchical framework in emerging mobility systems \citep{Bang2023flowbased}, and introduced a heuristic approach for low-level control. In this paper, we advance this approach by demonstrating that our low-level control strategies can attain optimality under specific conditions. We offer a comprehensive analysis of these conditions, thus enhancing our understanding of the circumstances conducive to optimal performance.
Moreover, when the solutions generated by our proposed method are inapplicable, our framework offers adaptive solutions to CAV trajectories. This adaptation naturally changes the travel time of CAVs.
Consequently, we provide a method of adjusting the traffic flow so that our solution approach regains feasibility. This flow modification allows us to utilize optimal trajectories at low-level controls while alleviating the computational complexities associated with solving constrained optimization problems.

We believe that the proposed hierarchical decision-making framework represents a pioneering approach to addressing the challenges of the emerging mobility landscape. It encompasses practical implementation, optimality conditions, and adaptive adjustments that collectively enhance the effectiveness and feasibility of decision-making processes in emerging mobility systems.

The structure of the paper is organized as follows.
In Section \ref{sec:flow_optimization}, we present the traffic flow optimization in a macroscopic perspective and the route-recovery algorithm at a microscopic level.
In Section \ref{sec:coordination}, we provide the coordination framework at an intersection.
In Section \ref{sec:trajectory_planning}, we present the trajectory planning method using Hamiltonian analysis,
and in Section \ref{sec:simulation}, we present simulation results to evaluate the performance of the proposed framework.
Finally, in Section \ref{sec:conclusion}, we draw concluding remarks and discuss future research directions.


\section{On-Demand Routing Framework}   \label{sec:flow_optimization}

In this section, we introduce a routing methodology for AMoD systems where all the vehicles are CAVs.
To utilize the routes at lower levels, we need to obtain a route for each CAV. In our framework, instead of solving the vehicle routing problem for each CAV, we solve the flow-based routing problem.
Typically, vehicle flow is assessed by the number of vehicles passing through a specific point within a given time frame, and the traffic flow is optimized from a macroscopic perspective without considering microscopic phenomena.
This allows us to obtain the system-wide optimal solution without a computational burden in a large-scale AMoD system.
Thus, we optimize the traffic flow at a macroscopic level and obtain a route for each CAV by using a heuristic route-recovery algorithm. Then, we provide rules for CAVs to depart at depots to generate corresponding flow.

\subsection{On-Demand Mobility Flow Optimization}

\begin{figure}
    \centering
    \includegraphics[width=0.9\linewidth]{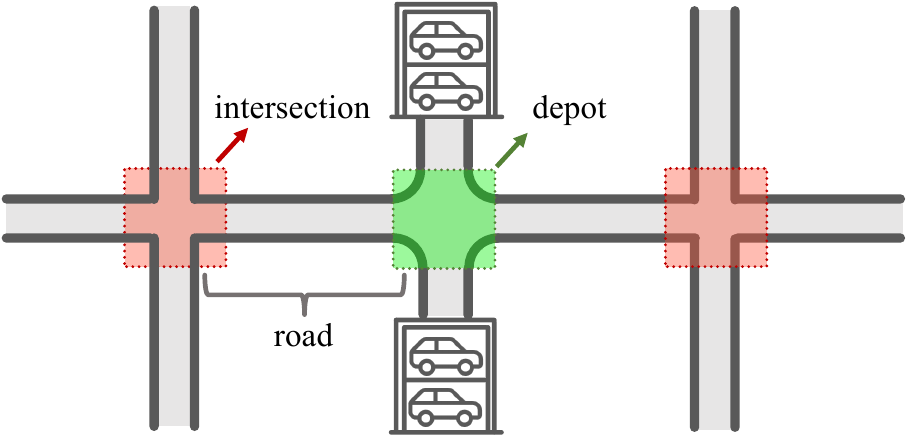}
    \caption{Illustration of the road network. The green and red squares represent depot and intersection nodes, respectively, and the road connecting those nodes represents the edge.}
    \label{fig:network_concept}
\end{figure}
We start our exposition by considering an urban grid road network represented as a directed graph $\mathcal{G} = (\mathcal{V},\mathcal{E}$), where $\mathcal{V}$ comprises vertices and $\mathcal{E}$ consists of edges connecting those vertices.
The vertices can be either intersections $r\in\mathcal{R}$ or depots $q\in\mathcal{D}$. Here, $\mathcal{R}$ and $\mathcal{D}$ denote the sets of intersections and depots, respectively, such that $\mathcal{V} = \mathcal{R} \cup \mathcal{D}$.
As illustrated in Fig. \ref{fig:network_concept}, depots are located between the exit of one intersection and the entry of its adjacent intersection.
This placement implies that a CAV passing through a depot transitions from one intersection to another. Next, we determine travel information.

Let $\mathcal{M}\in\mathbb{N}$ denote the total number of travel demands.
For each mobility demand $m\in\mathcal{M}$, we have an origin $o^m\in\mathcal{V}$, destination $d^m\in\mathcal{V}$, and demand rate $\alpha^m\in\mathbb{R}_{\geq0}$. The demand rate signifies the number of travelers per unit of time.
The flow of CAVs traveling on the road $(i,j)$ for each mobility demand $m$ is represented as $x_{ij}^m$. The total flow on road $(i,j)$ can be expressed as:
\begin{equation}
x_{ij} = \sum_{m\in\mathcal{M}} x_{ij}^m, \quad \forall (i,j)\in\mathcal{E}.
\end{equation}
To meet all the mobility demands, the flow must adhere to the following constraints:
\begin{align}
    &\sum_{i:(i,j)\in\mathcal{E}}x_{ij}^m = \sum_{k:(j,k)\in\mathcal{E}}x_{jk}^m,~\forall m\in\mathcal{M},j\in\mathcal{V}\setminus\{o^m,d^m\},\label{eqn:con_x1}\\
    & \sum_{k:(j,k)\in\mathcal{E}}x_{jk}^m = \alpha^m,~\forall m\in\mathcal{M},j=o^m,\label{eqn:con_x2}\\
    & \sum_{i:(i,j)\in\mathcal{E}}x_{ij}^m = \alpha^m,~\forall m\in\mathcal{M},j=d^m.\label{eqn:con_x3}
\end{align}
Constraint \eqref{eqn:con_x1} guarantees the flow conservation, i.e., the incoming and outgoing flows are equal at each node $j\in\mathcal{V}\setminus\{o^m,d^m\}$. Constraints \eqref{eqn:con_x2} and \eqref{eqn:con_x3} ensure CAVs to depart from origins and arrive to destinations with demand rate $\alpha^m$, respectively.

The travel time on the roads can be captured by the latency function suggested by the \textit{U.S. Bureau of Public Roads} (BPR) \citep{us1964traffic}.
By utilizing this function, we can express the travel time on road $(i,j)\in\mathcal{E}$ as
\begin{equation}
    t_{ij}(x_{ij}) = t_{ij}^0\left(1+0.15\left(\frac{x_{ij}}{\gamma_{ij}}\right)^4\right). \label{eqn:BPR}
\end{equation}
Here, $t_{ij}^0 \in\mathbb{R}_{>0}$ is the free-flow travel time and $\gamma_{ij} \in\mathbb{R}_{>0}$ is capacity of the road $(i,j)\in\mathcal{E}$.
Next, we introduce a flow optimization problem using the BPR function.

\begin{problem}[Flow-based routing] \label{prb:flow_routing}
The objective is to find the system-optimal flow of CAVs by solving the following optimization problem:
\begin{equation}
\begin{aligned}
    \min_{\mathbf{x}} ~&J(\mathbf{x}) = \sum_{(i,j)\in\mathcal{E}} \bigg\{ t_{ij}(x_{ij})x_{ij}\bigg\} \label{eqn:flow_routing}\\
    \text{s.t. } & \eqref{eqn:con_x1} \text{--} \eqref{eqn:con_x3}, \mathbf{x} \geq 0.
\end{aligned}    
\end{equation}
Here, $\mathbf{x}$ is a vector of flow $x_{ij}$ for all $(i,j)\in\mathcal{E}$.
\end{problem}

\begin{proposition}
Given the BPR function \eqref{eqn:BPR}, Problem \ref{prb:flow_routing} is a convex optimization problem.
\end{proposition}
\begin{proof}
Convexity is preserved through summation, so we need to demonstrate that $t_{ij}(x_{ij})x_{ij}$ is convex. Consider the function $f:\mathbb{R}_{\geq0} \to \mathbb{R}_{\geq0}$, given by $f(x_{ij}):=t_{ij}^0(1+0.15(x_{ij}/\gamma_{ij})^4)x_{ij}$. This is a fifth-order polynomial function, which is twice differentiable. We find that $\nabla^2 f(x_{ij})=0.15*20*t_{ij}^0(x_{ij}/\gamma_{ij})^3 \geq 0,~\forall x_{ij} \geq 0$.
Moreover, it is strictly increasing because $\nabla f(x_{ij}) \geq 0$, and $\nabla f(x_{ij}) = 0$ only if $x_{ij}=0$. Therefore, following the second-order derivative condition, $f(x_{ij})$ is strictly convex.
\end{proof}

Problem \ref{prb:flow_routing} aims to minimize the total travel time for all CAV flows.
Since Problem \ref{prb:flow_routing} is a strictly convex problem, it guarantees a unique global optimal solution.

\begin{remark}
    Generally, the free-flow travel time $t_{ij}^0$ is related to the geometry of the roads, such as the number of lanes, length, and capacity.
    Meanwhile, in our framework, we can utilize it as a base travel time that accounts for the traffic conditions at a microscopic level.
    This allows us to find new optimal flows for the actual traffic conditions when CAVs cannot implement the previously selected flows (see Section \ref{subsec:flow_modification}).
\end{remark}

\begin{remark}
In Problem \ref{prb:flow_routing}, we consider a 100\% penetration rate of CAVs in the network, which results in the travel time only depending on CAV flows.
This is because our focus is on utilizing the flow to obtain routes for CAVs. However, it is worth noting that we can extend the model to consider other scenarios such as relocating trips \citep{salazar2019congestion}, mixed traffic for CAVs \citep{wollenstein2020congestion,wollenstein2021routing}, or charging constraints \citep{bang2021AEMoD}, which would lead to non-convex problems.
In such cases, convexification can be achieved using piece-wise affine functions to approximate the BPR function.
\end{remark}

\subsection{Route Recovery and Departure Rules} \label{subsec:route_recovery}

The solution to Problem \ref{prb:flow_routing} determines the optimal flow on roads for each mobility demand $m\in\mathcal{M}$.
The optimal flow connecting from origin to destination can have multiple routes for CAVs.
Therefore, it is necessary to recover all different routes and corresponding flows to assign a specific route and departure time to each CAV.
Let $\mathcal{P}^m$ denote a set of routes for mobility demand $m\in\mathcal{M}$, where $|\mathcal{P}^m|\in\mathbb{N}$ corresponds to the total number of routes.
Each route $\mathcal{P}^m_l\in\mathcal{P}^m$, $l\in\{1,\dots,|\mathcal{P}^m|\}$, consists of a sequence of roads $(i,j)\in\mathcal{E}$ connected from $o^m$ to $d^m$, with $f_l^m$ representing the corresponding flow.

We recover all routes based on the optimal flow using Algorithm \ref{Alg:route_recovery}.
The primary concept underlying this algorithm is to search for connected flows from origin to destination.
This search process starts from the origin, finds the smallest flow among the connected road segments, and repeats until it reaches the destination. Then, we subtract the corresponding flow from the total demand rate and repeat the process until all the routes are recovered.
We prioritize exploring the most direct path first to minimize the number of turns for CAVs.

\begin{algorithm}[t]
 \caption{Heuristic Route Recovery}

\begin{algorithmic}[1] 
\For{$m\in\mathcal{M}$}
    \State{$l\gets 0$}
    \While{\texttt{not done}}
        \State{$l\gets l+1$;~~$i\gets o^m$;~~$f^m_l \gets$ $\alpha^m$}
        \\
        \State{\texttt{//Finding one possible path}}
        \While{$i \neq d^m$}
            \For{$j:(i,j)\in\mathcal{E}$}
                \If{ $x_{ij} \neq 0 $}
                    \State{$f^m_l \gets \min\{f^m_l,x_{ij}\}$}
                    \State{$\mathcal{P}^m_l \gets$ Add road $(i,j)$}
                    \State{$i \gets j$}
                    \State{\texttt{Break}}
                \EndIf
            \EndFor
        \EndWhile
        \\
        \State{\texttt{//Adjusting the remaining flow}}
        \For{$(i,j)\in\mathcal{P}^m_l$}
            \State{$x_{ij} \gets x_{ij} - f^m_l$}
        \EndFor
        \If{$\sum_{(i,j)}x_{ij} = 0$}
            \State{$|\mathcal{P}^m| \gets l$}
            \State{\texttt{done} $\gets$ \texttt{True}}
        \EndIf
    \EndWhile
\EndFor
\end{algorithmic} \label{Alg:route_recovery}
\end{algorithm}

\begin{figure}
    \centering
    \includegraphics[width=0.65\linewidth]{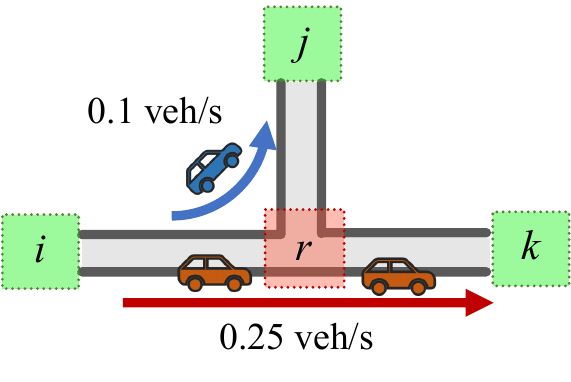}
    \caption{Illustration of CAVs departing at depot node $i\in\mathcal{D}$. The blue and red CAV flows follow the routes ($i\to r \to j$) and ($i\to r \to k$), respectively.}
    \label{fig:departure_concept}
\end{figure}

\begin{figure}
    \centering
    \subfloat[Asynchronous departure timing for each route]{\includegraphics[width=\linewidth]{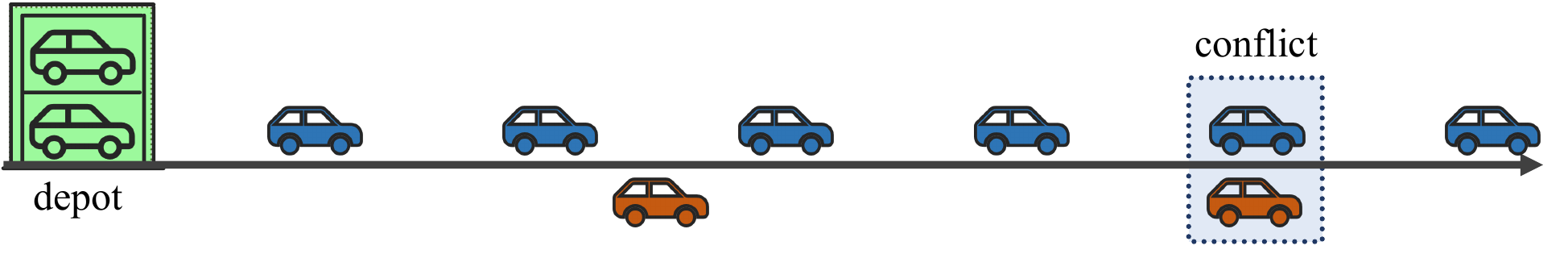}\label{b}}\\
    \subfloat[Synchronous departure timing at a depot]{\includegraphics[width=\linewidth]{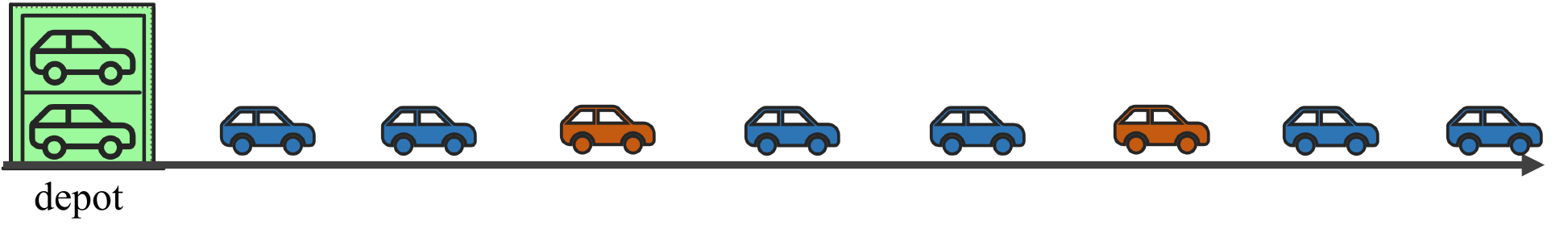}\label{c}}
    \caption{Illustration of departure time of CAVs for different flows. CAVs depart in the order of the illustrated queue.}
    \label{fig:departure_timing}
\end{figure}

Next, we need to establish rules for CAVs departing from the origins of the trips.
For instance, suppose we have obtained optimal flows for two different paths, illustrated in Fig. \ref{fig:departure_concept}. One group of CAVs (blue-colored flow in Fig. \ref{fig:departure_concept}) generates a flow of $0.1$ vehicles per second, while the other (red-colored flow) forms a flow of $0.25$ vehicles per second. This means that CAVs should enter the road $(i,r)$ and exit the road $(r,j)$ every $10$ seconds, and the roads $(i,r)$ and $(r,k)$ every $4$ seconds.
However, without synchronization, two CAVs from different flows may have conflict in the departure time as seen in Fig. \ref{fig:departure_timing}-(a), because each departure frequency does not account for CAVs in other paths, although they share depot $i$.
To overcome this challenge, we synchronize all the flows and allow CAVs to depart at a uniform frequency based on the optimal flow at the road $(i,r)\in\mathcal{E}$, while maintaining the order of actual departure timing.

Next, given the routes and departure time at origins, we determine the travel time of each CAV $n$ crossing an intersection.
Let $t_n^0$ be the entry time of CAV $n\in\mathcal{N}$ at an intersection. Let $(i,r)\in\mathcal{E}$ and $(r,k)\in\mathcal{E}$ be the edges of the entry road and exit road of the intersection, respectively.
We select the exit time based on the following condition:
\begin{equation}
    t_n^f = \max\left\{t_n^0+t_{ir}(x^*_{ir})+t_{rk}(x^*_{rk}),~t_p^f+\frac{1}{x^*_{rk}}\right\}. \label{eqn:terminal_time}
\end{equation}
Here, $t_p^f$ is the exit time of preceding CAV $p$ that exits just before CAV $n$, and $x^*_{(\cdot)}$ is the optimal flow on the given road segment.
We choose the maximum value between two options in \eqref{eqn:terminal_time}.
The first option is the estimated arrival time at the exit node, utilizing the BPR function.
If no preceding CAV exits from the same node, CAV $n$ follows the estimated travel time.
However, the entering frequencies for different intersection entries are not synchronized.
This could lead to the existence of CAV $p$ entering at a different entry but having an overlapping exit time with CAV $n$.
To address this, we allow CAV $n$ to select the second option in \eqref{eqn:terminal_time} to exit after CAV $p$, while still maintaining the optimal frequency of the road $(r,k)$.

By using the aforementioned rules, CAVs can effectively determine their departure and arrival times at each intersection along their route.
Nevertheless, we have yet to explore the practicality of achieving these specified departure and arrival times with the actual speed trajectories of CAVs.
In the following section, we delve deeper into the coordination challenges at intersections, presenting a methodology to address this aspect.

\section{Flow-Matching Coordination Problem} \label{sec:coordination}

\begin{figure}
    \centering
    \includegraphics[width=0.7\linewidth]{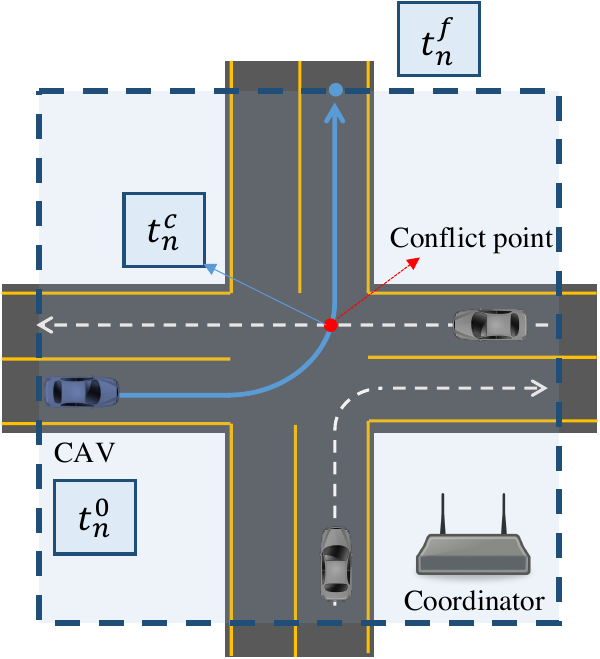}
    \caption{Coordination of CAVs at a signal-free intersection.}
    \label{fig:intersection}
\end{figure}

In this section, we introduce a coordination problem at a signal-free intersection.
The problem addresses energy-optimal trajectory planning for all CAVs with entry and exit times, which are specified from the previous section, to maintain the optimal flow.
We consider a single-lane intersection as illustrated in Fig. \ref{fig:intersection}, which comprises four entries and four exits.
Excluding U-turns from consideration, the intersection accommodates a total of twelve possible paths for CAVs.
Each intersection, denoted by $r\in\mathcal{R}$, is equipped with a \textit{coordinator} that communicates essential information with CAVs, including geometry details and trajectories of other CAVs.
Upon entering the intersection, CAVs strategically plan their trajectories in the sequence of entrance considering the presence of other CAVs within the intersection.
There is a set of points $\mathcal{C}\subset\mathbb{N}$ called \textit{conflict points}, which identifies locations where potential lateral collisions might occur among the CAVs.

Let $\mathcal{N}_r(t)$ be the set of CAVs in the intersection $r\in\mathcal{R}$ at time $t\in\mathbb{R}_{\geq 0}$.
The dynamics of each CAV $n\in\mathcal{N}_r(t)$ are given by double integrator, i.e.,
\begin{equation}
    \begin{aligned}
    \dot{s}_n(t) &= v_n(t),\\
    \dot{v}_n(t) &= u_n(t), \label{eqn:dynamics}
    \end{aligned}
\end{equation}
where $s_n(t)\in\mathbb{R}_{\geq0}$ represents the distance from the entry point of the intersection to the current location of CAV $n$, $v_n(t)\in\mathbb{R}_{\geq0}$ denotes the speed, and $u_n(t)\in\mathbb{R}$ is the control input of CAV $n$.
For each CAV $n\in\mathcal{N}_r(t)$, state and input constraints are imposed to ensure safe operation:
\begin{align}
    u_{n,\text{min}} & \leq u_n(t) \leq u_{n,\text{max}}, \label{eqn:ulim}\\
    0 < v_{\text{min}} & \leq v_n(t) \leq v_{\text{max}}, \label{eqn:vlim}
\end{align}
where $u_{n,\text{min}}$, $u_{n,\text{max}}$ represent the minimum and maximum control inputs of CAV $n$ and $v_{\text{min}}$, $v_{\text{max}}$ are the minimum and maximum allowable speeds at the intersection, respectively.

We enforce two types of constraints to ensure safety: rear-end safety and lateral safety.
We impose rear-end safety that requires a minimum distance $\delta \in\mathbb{R}_{>0}$ between preceding CAV $p\in\mathcal{N}_r(t)$ and following CAV $n\in\mathcal{N}_r(t)$, i.e.,
\begin{equation}
    s_p(t) - s_n(t) \geq \delta. \label{eqn:rear-end}
\end{equation}
For lateral safety, consider CAV $p$ is on a different path with CAV $n$ and shares conflict point $c\in\mathcal{C}$.
Then, we ensure lateral safety by imposing constraints on the time separation between CAVs $p$ and $n$ at conflict points $c$:
\begin{equation}
    |t_p^c - t_n^c | \geq \tau^\mathrm{safe}. \label{eqn:lateral}
\end{equation}
Here, $\tau^\mathrm{safe}\in\mathbb{R}_{\geq0}$ represents the safety time headway, and $t_p^c$ and $t_n^c$ denote the arrival times of CAVs $p$ and $n$ at conflict point $c$, respectively.

In our coordination framework, the objective is to minimize speed transitions of CAVs, thereby conserving momentum and reducing fuel consumption.
\begin{problem}
To find the minimum-energy trajectory, each CAV $n\in\mathcal{N}_r(t)$ solves the following optimal control problem
\begin{align} \label{eqn:energy-optimal}
    \min &\text{~} \frac{1}{2} \int_{t_n^0}^{t_n^f} u_n^2(t) dt\\
    \emph{s.t. }& \eqref{eqn:dynamics} -  \eqref{eqn:lateral} \notag,
\end{align}
where $t_n^0$ is the entry time and $t_n^f$ is the exit time of CAV $n$ at the intersection, respectively.
    \label{prb:constrained_energy_optimal}
\end{problem}

Due to the safety constraints \eqref{eqn:rear-end} and \eqref{eqn:lateral}, it is challenging to find an analytical solution to Problem \ref{prb:constrained_energy_optimal}.
Even considering only state/input constraints, it is required to piece constrained and unconstrained arcs together to find the optimal trajectory, which is inapplicable for real-time implementation \citep{malikopoulos2019ACC,Malikopoulos2020}.
Our previous work \citep{Malikopoulos2019CDC,Malikopoulos2020} addressed a similar problem and presented a method of using unconstrained energy-optimal trajectory and adjusting the exit time not to violate any constraints.
However, this method cannot be applied in our proposed framework since changing the exit time will directly affect the flow and reduce the efficiency of the selected routes.
Considering these challenges, we provide a different method in the next section, which can solve the problem efficiently and expand the set of feasible solutions to accommodate the flow constraints better.


\section{Hamiltonian-Based Trajectory Planning} \label{sec:trajectory_planning}

Given the double-integrator dynamics for each CAV $n\in\mathcal{N}$, the Hamiltonian with state and control input constraints is given as
\begin{equation}
    \begin{aligned} \label{eqn:hamiltonian_full}
        \mathcal{H}_n(&s_n(t),v_n(t),u_n(t),t)\\
         & = \frac{1}{2} u_n^2(t) + \lambda^s_n v_n(t) + \lambda^v_n u_n(t)\\
        & + \mu^a_n (v_n(t)-v_{\max} ) + \mu^b_n (v_{\min} - v_n(t)) \\
        & + \mu^c_n (u_n(t)-u_{\max}) + \mu^d_n (u_{\min} - u_n(t)),
    \end{aligned}
\end{equation}
where $\lambda^s_n$, $\lambda^v_n$ are costates and $\mu^a_n,\mu^b_n,\mu^c_n,\mu^d_n$ are Lagrange multipliers.
From Pontryagin's minimum principle, the Euler-Lagrange equations become
\begin{equation}
\label{eqn:lagrange}
    \begin{aligned}
        \dot{\lambda}^s_n = -\frac{\partial \mathcal{H}_n}{\partial s_n} &= 0, \\
        \dot{\lambda}^v_n = -\frac{\partial \mathcal{H}_n}{\partial v_n} &= \lambda^s_n+\mu^a_n-\mu^b_n, \\
        0 = \frac{\partial \mathcal{H}_n}{\partial u_n} &= u_n+\lambda^v_n+\mu^c_n-\mu^d_n.
    \end{aligned}
\end{equation}


Next, consider the case of no state and control input constraints become active, i.e., $\mu^a_n=\mu^b_n=\mu^c_n=\mu^d_n=0$.
Then, from \eqref{eqn:lagrange}, the Lagrange multipliers and control input must satisfy 
\begin{equation}
\begin{aligned}
    \dot{\lambda}^s_n(t) &= 0, \\
    \dot{\lambda}^v_n(t) &= \lambda^s_n, \\
    u_n(t) &= -\lambda^v_n. \label{eqn:lambda}
\end{aligned}    
\end{equation}
Since $\lambda^s_n(t)$ is constant, $u_n(t)=-\lambda^v_n$ becomes linear function, i.e., $u_n(t) = 6a_nt+2b_n$.
Consequently, the optimal trajectory has the following form:
\begin{align} \label{eq:optimalTrajectory}
    u_n(t) &= 6 a_n t + 2 b_n, \notag \\
    v_n(t) &= 3 a_n t^2 + 2 b_n t + c_n, \\
    s_n(t) &= a_n t^3 + b_n t^2 + c_n t + d_n, \notag
\end{align}
where $a_n, b_n, c_n$, and $d_n$ are constants of integration.
These constants can be computed from the boundary conditions
\begin{align}
     s_n(t_n^0) &= 0,\quad  v_n(t_n^0)= v_n^0 , \label{eq:bci}\\
     s_n(t_n^f)&=s_n^f,\quad v_n(t_n^f)=v_n^f. \label{eq:bcf}
\end{align}
Here, $s_n^f$ is the road length from the entry to the exit, $v_n^0$ is the entry speed, and $v_n^f$ is the exit speed of CAV $n$.
Given the boundary conditions \eqref{eq:bci} and \eqref{eq:bcf}, we can easily compute the constants of integration and obtain the unconstrained energy-optimal trajectory \eqref{eq:optimalTrajectory}.
\begin{remark}
The values of $v_n^0$ and $v_n^f$ can be determined based on the estimated travel time given from the flow optimization. For example, let $t_0^\textrm{BPR}$ and $t_f^\textrm{BPR}$ denote the estimated travel time resulting from \eqref{eqn:terminal_time} for the entry and exit roads, respectively. Then, the entry speed and exit speed can be determined by
\begin{equation}
    v_n^0 = \frac{s_n^f}{2t_0^\mathrm{BPR}},\quad v_n^f = \frac{s_n^f}{2t_f^\mathrm{BPR}}. \label{eqn:speed_cond}
\end{equation}
This provides reliable $v_n^0$ and $v_n^f$ since traveling at $v_n^0$ for the half of the road and $v_n^f$ for the other half naturally takes traveling time $t_n^f$ given by \eqref{eqn:terminal_time}.
\end{remark}

Generally, it is necessary to numerically solve a constrained optimization problem if the unconstrained trajectory violates any of the constraints.
In the following subsection, as one of our main contributions, we provide a method that utilizes the form of unconstrained trajectories to compute analytical solutions for each possible violation.
This method will allow us to avoid the computational burden of solving the constrained problem numerically and obtain the optimal trajectory in real time.

\begin{remark}
    Note that the unconstrained energy-optimal trajectory is polynomial; thus, it is straightforward to find extrema and verify whether the state/input constraints are violated.
    As we have fixed the boundary conditions (e.g., $t_n^0,t_n^f,v_n^0,v_n^f$), if the unconstrained energy-optimal trajectory violates state/input constraints, the feasible solution may not exist for the given boundary conditions.
    Since the boundary conditions result from the flow optimization, it is necessary to modify the flow in such cases.
    The method will be explained in Section \ref{subsec:flow_modification}.
\end{remark}


\subsection{Case of Lateral Safety Violation}
\begin{figure}
    \centering
    \includegraphics[width=\linewidth]{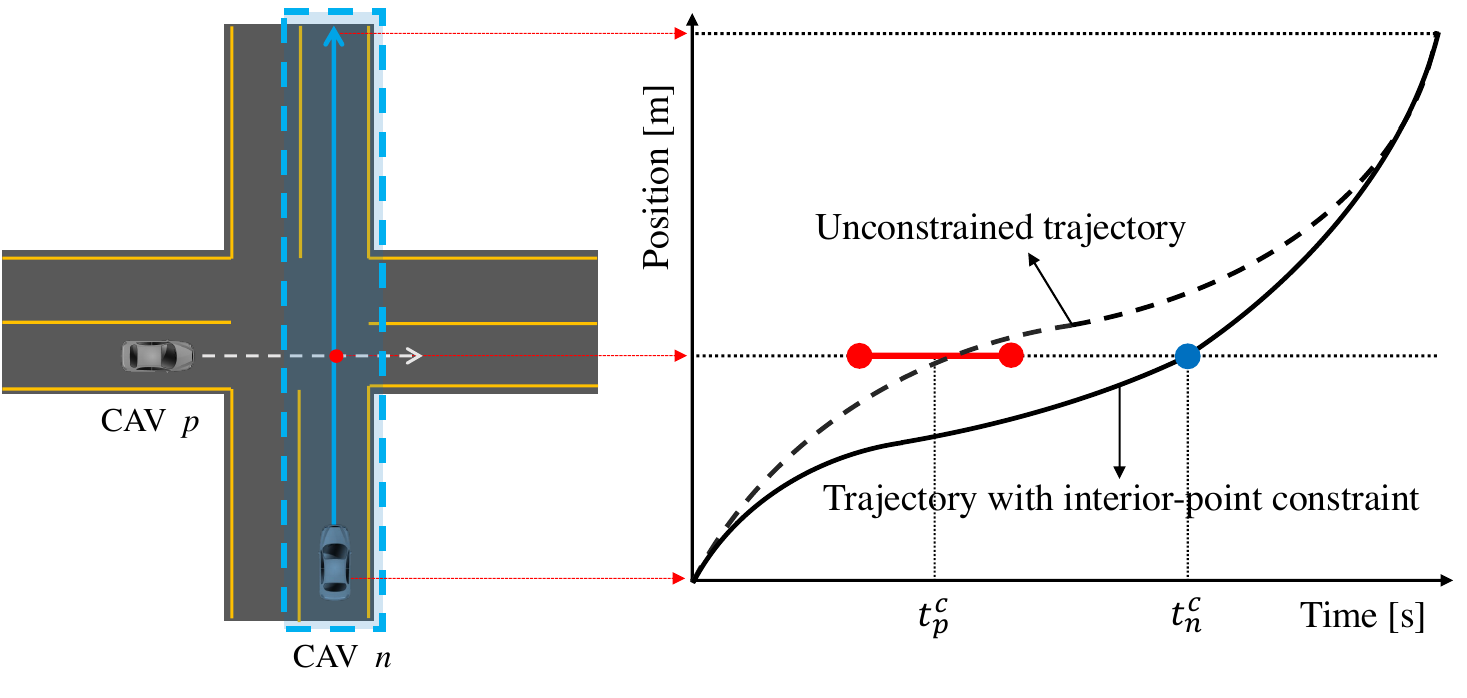}
    \caption{Schematic diagram of lateral safety violation. Blue dot represents the interior point constraint.}
    \label{fig:lateral}
\end{figure}

If the unconstrained trajectory \eqref{eq:optimalTrajectory} fails to meet the lateral safety requirement specified in constraint \eqref{eqn:lateral}, the optimization problem must incorporate this constraint.
However, as previously highlighted, addressing the constrained problem involves seamlessly integrating both constrained and unconstrained arcs, yielding a computational challenge for real-time implementation.
To resolve this challenge, we introduce a novel approach utilizing an interior-point constraint (see Fig. \ref{fig:lateral}).
Leveraging this interior-point constraint, we can derive an optimal trajectory that navigates through a specific point. Subsequently, we identify the best interior point that simultaneously optimizes the objective function in Problem \ref{prb:constrained_energy_optimal} while satisfying the lateral safety constraints.
This approach is highly effective, as it eliminates the need for complicated numerical computations and enables the rapid acquisition of an optimal solution.

The interior-point constraint denoted by $N(x_n(t),t) := s_n(t)-s_n^c$ must satisfy $N(x_n(t_n^c),t_n^c) = 0$ where $t_n^c$ is the time of a CAV passing the point $s_n^c$.
Let ${t_n^c}^-$ and ${t_n^c}^+$ signify just before and after $t_n^c$, respectively.
Then, the following jumping conditions are necessary for the solution trajectory to be optimal [Ch 3.5, \cite{bryson1975applied}]:
\begin{align}
    \lambda^s_n ({t_n^c}^-) & = \lambda^s_n ({t_n^c}^+) + \pi^s_n \label{eqn:lambda_s}\\
    \lambda^v_n ({t_n^c}^-) & = \lambda^v_n ({t_n^c}^+), \label{eqn:lambda_v}\\
    \mathcal{H}_n({t_n^c}^-) & = \mathcal{H}_n({t_n^c}^+).
\end{align}

The above conditions along with \eqref{eqn:lambda} imply that $\lambda^s_n(t)$ is discontinuous at $t_n^c$ while $\lambda^v_n(t)$ is continuous. Specifically, $u_n(t)$ will be continuous in all $t\in[t_n^0,t_n^f]$ but with different coefficients for $t\in[t_n^0,t_n^c]$ and $t\in[t_n^c,t_n^f]$ due to \eqref{eqn:lambda_s}.
Imposing the interior point constraints, the optimal solution has the following structural form 
\begin{align}\label{eqn:optimal_u}
    u_n(t) = \begin{dcases}
        6a_1t+2b_1, & \text{if~~} {t_n^0} \leq t < {t_n^c},\\
        6a_2t+2b_2, & \text{if~~} {t_n^c} \leq t \leq {t_n^f},
        \end{dcases}
\end{align}
\begin{align}\label{eqn:optimal_v}
    v_n(t) = \begin{dcases}
        3a_1t^2+2b_1t+c_1, & \text{if~~} {t_n^0} \leq t < {t_n^c},\\
        3a_2t^2+2b_2t+c_2, & \text{if~~} {t_n^c} \leq t \leq {t_n^f},
        \end{dcases}
\end{align}
\begin{align}\label{eqn:optimal_s}
    s_n(t) = \begin{dcases}
        a_1t^3+b_1t^2+c_1t+d_1, & \text{if~~} {t_n^0} \leq t < {t_n^c},\\
        a_2t^3+b_2t^2+c_2t+d_2, & \text{if~~} {t_n^c} \leq t \leq {t_n^f},
        \end{dcases}
\end{align}
which are joint functions of two unconstrained energy-optimal trajectories.
Using the above trajectories, the objective function can be expressed by 
\begin{equation}
    J  = \frac{1}{2}\left\{ \int_{t_n^0}^{{t_n^c}} (6a_1t+2b_1)^2 dt + \int_{t_n^c}^{t_n^f} (6a_2t+2b_2)^2 dt \right \}.
\end{equation}

To determine all constants in \eqref{eqn:optimal_u}--\eqref{eqn:optimal_s}, we use the boundary conditions, i.e., $v_n(t_n^c)=v_n^c$, $s_n(t_n^c)=s_n^c$, \eqref{eq:bci}, \eqref{eq:bcf}.
However, in the current formulation, the computation necessitates the inversion of a $4\times4$ matrix and solving a linear matrix equation to establish these constants. This not only imposes a heavier computational load on the onboard system but also complicates the problem, rendering it challenging for analytical comprehension.

To address these issues, we shift time and distance so that the trajectories always start at $t=0$ and $s_n(0)=0$.
We introduce tilde notation to distinguish constants of integration for shifted-time coordination, i.e., $\Tilde{u}_n(t)=6\Tilde{a}_1t+2\Tilde{b}_1$ for $t\in[0,t_n^c-t_n^0]$.
This approach directly determines $\Tilde{c}_1 = v_n^0$, $\Tilde{c}_2 = v_n^c$, and  $\Tilde{d}_1 = \Tilde{d}_2 = 0$, and makes the problem more interpretable.
The remaining constants can be computed using the following equations:
\begin{equation}
    \left[\begin{matrix}\Tilde{a}_1\\\Tilde{b}_1 \end{matrix} \right] = \left[\begin{matrix}3\Delta {t_n^c}^2 & 2\Delta {t_n^c} \\\Delta {t_n^c}^3 &\Delta {t_n^c}^2 \end{matrix}\right]^{-1} \left[\begin{matrix}{v_n^c}-{v_n^0} \\ {s_n^c}-{v_n^0}\Delta {t_n^c} \end{matrix} \right], \label{eqn:traj1}
\end{equation}
\begin{equation}
    \left[\begin{matrix}\Tilde{a}_2\\\Tilde{b}_2 \end{matrix} \right] = \left[\begin{matrix}3\Delta {t_n^f}^2 & 2\Delta {t_n^f} \\\Delta {t_n^f}^3 &\Delta {t_n^f}^2 \end{matrix}\right]^{-1} \left[\begin{matrix} {v_n^f}-{v_n^c}\\\Delta {s_n^f}-{v_n^c}\Delta {t_n^f} \end{matrix} \right], \label{eqn:traj2}
\end{equation}
where $\Delta {t_n^c} = {t_n^c}-{t_n^0}$, $\Delta {t_n^f} = {t_n^f}-{t_n^c}$, and $\Delta {s_n^f} = {s_n^f}-{s_n^c}$.

Given the above constants, we solve the following optimization problem to find energy-optimal time and speed for passing the conflict point ${s_n^c}$:
\begin{equation}
\begin{aligned}
    \min_{{t_n^c},{v_n^c}} &\frac{1}{2}\left\{ \int_{0}^{\Delta {t_n^c}} (6\Tilde{a}_1t+2\Tilde{b}_1)^2 dt + \int_{0}^{\Delta {t_n^f}} (6\Tilde{a}_2t+2\Tilde{b}_2)^2 dt \right \} \\
    \text{s.t. }& \text{\eqref{eqn:lateral}}.\label{eqn:lateral_optimal}
\end{aligned}
\end{equation}
Substituting \eqref{eqn:traj1} and \eqref{eqn:traj2} into the objective function of \eqref{eqn:lateral_optimal}, we obtain new objective function
\begin{equation}
\begin{aligned}
    J = \frac{2}{\Delta {t_n^c}^3} \Bigg\{ \Delta {t_n^c}^2 ({v_n^c}^2+{v_n^c}{v_n^0}+{v_n^0}^2) \\
    - 3({v_n^c}+{v_n^0})\Delta {t_n^c} {s_n^c} + 3{s_n^c}^2\Bigg \}\\
    + \frac{2}{\Delta {t_n^f}^3} \Bigg\{ \Delta {t_n^f}^2 ({v_n^f}^2+{v_n^f}{v_n^c}+{v_n^c}^2)\\
    - 3({v_n^f}+{v_n^c})\Delta {t_n^f} \Delta {s_n^f} + 3\Delta {s_n^f}^2\Bigg \}. \label{eqn:lateral_cost}
\end{aligned}
\end{equation}

Based on \eqref{eqn:lateral_cost}, we establish the following lemma and theorem.

\begin{lemma} \label{lem:v_n^c}
For $t_n^c\in(t_n^0,t_n^f)$, $J$ is convex with respect to $v_n^c$.
\end{lemma}

\begin{proof}
    By computing the second partial derivative of $J$ with respect to $v_n^c$, we get $\frac{\partial^2 J}{\partial {v_n^c}^2} = \frac{4}{\Delta{t_n^c}}+\frac{4}{\Delta{t_n^f}}$.
    For $t_n^c\in(t_n^0,t_n^f)$, it follows that $\Delta{t_n^c}>0$ and $\Delta{t_n^f}>0$.
    Consequently, $\frac{\partial^2 J}{\partial {v_n^c}^2} > 0$, which satisfies second order condition for convexity.
\end{proof}

\begin{corollary}
From Lemma \ref{lem:v_n^c}, there exists a unique optimal speed of $v_n^c$.
Thus, we obtain the optimal speed ${v_n^c}^*$ by using $\frac{\partial J}{\partial {v_n^c}}=0$, i.e.,
\begin{equation} \label{eqn:opt_vc}
    {v_n^c}^* = \frac{3({s_n^c}\Delta {t_n^f}^2 + \Delta {s_n^f} \Delta {t_n^c}^2) - \Delta {t_n^c} \Delta {t_n^f} ({v_n^0} \Delta {t_n^f} + {v_n^f} \Delta {t_n^c})}{2\Delta {t_n^c} \Delta {t_n^f} (\Delta {t_n^f} + \Delta {t_n^c})}.
\end{equation}
\end{corollary}

\begin{theorem} \label{th:optimality}
For a ${v_n^c}^*$ given in \eqref{eqn:opt_vc}, let $J$ have a single extrema with respect to $t^c\in(t_n^0,t_n^f)$.
Then, the optimal time ${t_n^c}^*$, which is the solution to \eqref{eqn:lateral_optimal}, must exist on the boundary of the constraint, i.e., $|t_p^c-{t_n^c}^*| = \tau^\mathrm{safe}$.
\end{theorem}

\begin{proof}
    Substituting \eqref{eqn:opt_vc} into \eqref{eqn:lateral_cost}, $J$ becomes a function of time $t$.
    Due to $\Delta {t_n^c}^3$ and $\Delta {t_n^f}^3$ terms in $J$, $\lim_{t_n^c \to {t_n^0}^+} J = \infty$ and $\lim_{t_n^c \to {t_n^f}^-} J = \infty$.
    That is, the extrema of $J$ becomes a minima.
    Note that the optimal time ${t_n^c}^*$ minimizing $J$ must lie on \eqref{eq:optimalTrajectory} because it is the optimal trajectory for the general unconstrained problem.
    Consequently, $\pi_n^s=0$ in \eqref{eqn:lambda_s}, and \eqref{eqn:optimal_u}-\eqref{eqn:optimal_s} would yield smooth polynomials for $t\in[t_n^0,t_n^f]$.
    Considering the case where the unconstrained energy-optimal trajectory violates the lateral safety constraint, crossing the conflict point at time ${t_n^c}^*$ would violate the constraint, i.e., $|t_p^c-{t_n^c}^*| < \tau^\mathrm{safe}$.
    Since $J$ is unimodal function, $J(t_1) > J(t_2)$ for $t_n^0<t_1<t_2<{t_n^c}^*$ and $J(t_3) < J(t_4)$ for ${t_n^c}^*<t_3<t_4<t_n^f$.
    Therefore, the minimum $J$ satisfying the constraint must be on the boundary of the constraint, i.e., $|t_p^c-{t_n^c}^*| = \tau^\mathrm{safe}$.
\end{proof}

Next, we find sufficient conditions for $J$ to be an unimodal function, which has a single extrema, and discuss the underlying physical meaning of the conditions.
Note that, if these conditions hold, we can easily obtain the optimal solution given by \eqref{eqn:optimal_u}--\eqref{eqn:optimal_s}, where ${t_n^c}^*=t_p^c-\tau$ or ${t_n^c}^*=t_p^c+\tau$.

After substituting \eqref{eqn:opt_vc} into \eqref{eqn:lateral_cost} and differentiating the result, we obtain
\begin{equation}
\begin{aligned} 
    \frac{dJ}{dt_n^c} = & \frac{2v_n^c}{\Delta {t_n^c}^2 \Delta {t_n^f}^2}\cdot \Big[ 2\Delta t_n^c(3\Delta s_n^f - v_n^f \Delta t_n^f)\\
    & - 2\Delta t_n^f(3 s_n^c - v_n^0 \Delta t_n^c)\\
    &- \left\{ \Delta {t_n^c}^2(v_n^f-v_n^c) + \Delta {t_n^f}^2 (v_n^c-v_n^0)\right\} \Big].
\end{aligned}
\end{equation}

Let $\Gamma := 2\Delta t_n^c(3\Delta s_n^f - v_n^f \Delta t_n^f) - 2\Delta t_n^f(3 s_n^c - v_n^0 \Delta t_n^c) - \left\{ \Delta {t_n^c}^2(v_n^f-v_n^c) + \Delta {t_n^f}^2 (v_n^c-v_n^0)\right\}$.
Due to \eqref{eqn:vlim}, $v_n^c$ should always be positive.
Therefore, $\Gamma$ must be unimodal with respect to $t_n^c$ in order $J$ to be unimodal, i.e., $\frac{d\Gamma}{d t_n^c} > 0$.
Differentiating $\Gamma$ with respect to time, we obtain
\begin{subequations}
    \begin{align}
    2\cdot \frac{d\Gamma}{dt_n^c} &= \nonumber\\ 
    & \left( 3+2\frac{\Delta t_n^c}{\Delta t_n^f} + \frac{\Delta {t_n^c}^2}{\Delta {t_n^f}^2}\right) ( 3\Delta s_n^f - v_n^f \Delta t_n^f) \label{eqn:sub1}\\
    +& \left( 3+2\frac{\Delta t_n^f}{\Delta t_n^c} + \frac{\Delta {t_n^f}^2}{\Delta {t_n^c}^2}\right) ( 3 s_n^c - v_n^0 \Delta t_n^c) \label{eqn:sub2}\\
    +&~ \frac{\Delta t_n^c - \Delta t_n^f}{\Delta t_n^c \Delta t_n^f} \left( \Delta {t_n^c}^2 v_n^f - \Delta {t_n^f}^2 v_n^0 \right).\label{eqn:sub3}
\end{align}
\label{eqn:diff_gamma}
\end{subequations}

Next, we construct sufficient conditions for ensuring $\frac{d\Gamma}{d t_n^c} > 0$.
First, we find conditions for the sum \eqref{eqn:sub1} $+$ \eqref{eqn:sub2} $>0$, as these two terms significantly influence determining the sign of $\frac{d\Gamma}{d t_n^c}$.
Let $T$ denote the total travel time to cross the intersection, i.e., $T=t_n^f-t_n^0=\Delta t_n^f + \Delta t_n^c$.
Then, there are two possible cases.
(i) When $\Delta t_n^c > \Delta t_n^f$,  we impose the condition that $( 3\Delta s_n^f - v_n^f \Delta t_n^f) > ( 3 s_n^c - v_n^0 \Delta t_n^c)$.
This condition ensures that the sum \eqref{eqn:sub1} $+$ \eqref{eqn:sub2} is positive, even in the worst-case scenarios of $\Delta t_n^c = T$ or $\Delta t_n^f=0$.
(ii) When $\Delta t_n^c \leq \Delta t_n^f$, we instead impose the condition $T<\frac{3s_n^f}{v_n^f}$ to ensure \eqref{eqn:sub1} $+$ \eqref{eqn:sub2} to be positive.
To avoid the case where both \eqref{eqn:sub1} and \eqref{eqn:sub2} become negative, we provide additional conditions forcing both terms to be positive when $\Delta t_n^c = \Delta t_n^f = \frac{T}{2}$. That is, $\frac{T}{2} < \frac{3s_n^c}{v_n^0}$ and $\frac{T}{2} < \frac{3\Delta s_n^f}{v_n^f}$.
Gathering all the conditions, we impose the following constraint to ensure \eqref{eqn:sub1} $+$ \eqref{eqn:sub2} $>0$:
\begin{equation}
    T < \min \left\{ \frac{3s_n^f}{v_n^0},\frac{3s_n^f}{v_n^f},\frac{6  s_n^c}{v_n^0},\frac{6\Delta s_n^f}{v_n^f} \right\}.\label{eqn:lateral_cond1}
\end{equation}

Next, we establish more conditions upon \eqref{eqn:lateral_cond1} to ensure \eqref{eqn:diff_gamma} to be positive.
Recalling that \eqref{eqn:sub1} and \eqref{eqn:sub2} are critical in determining the sign of $\frac{d\Gamma}{d t_n^c}$, $\frac{d\Gamma}{d t_n^c}$ is positive if \eqref{eqn:sub1} $>0$, \eqref{eqn:sub2} $>0$ and negative if \eqref{eqn:sub1} $<0$, \eqref{eqn:sub2} $<0$.
The only case of \eqref{eqn:sub3} influencing in the sign of $\frac{d\Gamma}{d t_n^c}$ is when either one of \eqref{eqn:sub1} and \eqref{eqn:sub2} is negative while the other is positive.
We use the following condition to ensure that \eqref{eqn:sub3} $>0$ in such a case.

\begin{equation}
    T < \min \left\{\frac{3 \Delta s_n^f}{v_n^f} \cdot\left( 1+ \sqrt{\frac{v_n^0}{v_n^f}}\right),\frac{3 s_n^c}{v_n^0} \cdot\left( 1+ \sqrt{\frac{v_n^f}{v_n^0}}\right)\right\}.\label{eqn:lateral_cond2}
\end{equation}
This condition can be derived using similar steps for \eqref{eqn:lateral_cond1}.
As a result, \eqref{eqn:sub3} remains positive when either \eqref{eqn:sub1} or \eqref{eqn:sub2} is negative.

If the total travel time $T$ satisfies both conditions \eqref{eqn:lateral_cond1} and \eqref{eqn:lateral_cond2}, $J$ becomes an unimodal function and thus, the optimal solution becomes \eqref{eqn:optimal_u}--\eqref{eqn:optimal_s} with $t_n^c=t_p^c-\tau^\mathrm{safe}$ or $t_n^c=t_p^c+\tau^\mathrm{safe}$ as in Theorem \ref{th:optimality}.

\begin{remark}
Conditions \eqref{eqn:lateral_cond1} and \eqref{eqn:lateral_cond2}  indicate that the total travel time requires an upper bound, which purely depends on the results of the traffic flow optimization and the route recovery process.
From a physical perspective, these conditions imply that the actual travel time $T$ must be smaller than the expected travel time with some gap.
Considering \eqref{eqn:terminal_time}, these conditions must hold unless the optimal flow goes significantly over the road capacity.
Nevertheless, if the conditions do not hold, it is necessary to modify the flow since the actual travel time deviates too much from the expected travel time, which would also fail to generate the optimal flows.
\end{remark}

\subsection{Case of Rear-End Safety Violation}

This subsection analyzes specific scenarios where rear-end safety violations might occur.
The primary scenario involves two CAVs, preceding CAV $p$ and following CAV $n$, traveling in the same path. Consider the situation where preceding CAV $p$ follows an unconstrained trajectory, which implies that CAV $p$ travels without any interruptions or impediments caused by other CAVs.
Thus, if the following CAV $n$ violates rear-end safety with the unconstrained energy-optimal trajectory, it suggests that the current traffic flow on the path is bringing CAVs too close, causing a chain reaction of disruptions for subsequent CAVs.

In such a case, to satisfy the rear-end safety constraints, all CAVs need to deviate from their unconstrained energy-optimal path and readjust their trajectory by incorporating additional deceleration and acceleration.
Implementing this extra deceleration and acceleration across all CAVs would demand a significantly higher energy cost. Hence, adjusting the traffic flow is more effective in this scenario so CAVs can follow an unconstrained energy-optimal trajectory. Section \ref{subsec:flow_modification} provides details on how to modify the flow.

\begin{figure}
    \centering
    \includegraphics[width=0.9\linewidth]{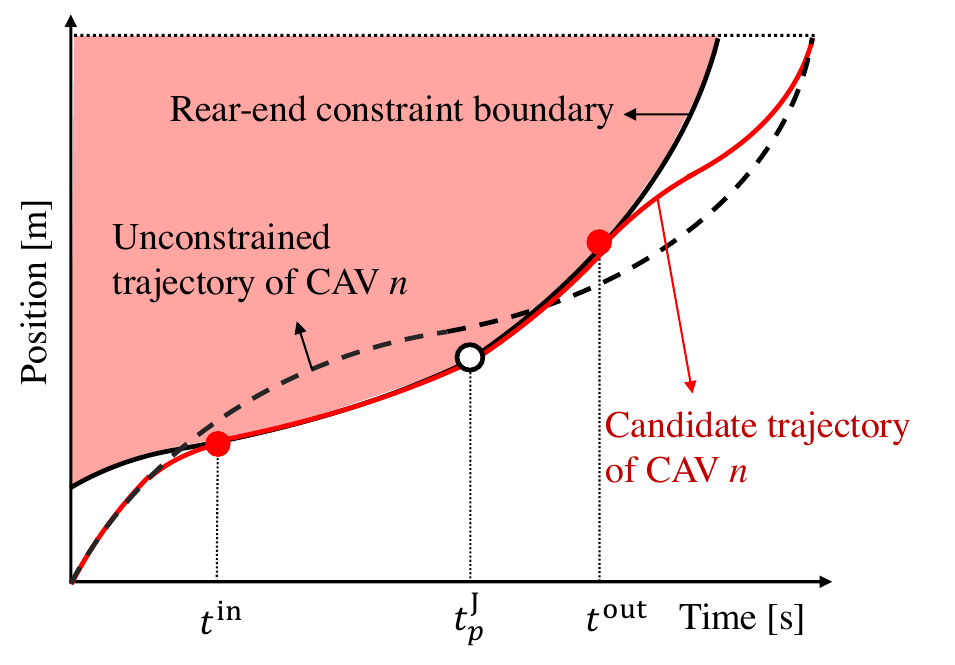}
    \caption{Schematic diagram of rear-end safety violation for CAV $n$. Candidate trajectory smoothly enters and exits the constraint boundary at $t^\mathrm{in}$ and $t^\mathrm{out}$, respectively.}
    \label{fig:rear_end_violation_case}
\end{figure}
Meanwhile, consider a scenario where the preceding CAV $p$ follows a constrained trajectory resulting from lateral safety violation with a junction point $(t_p^\mathrm{J},s_p^\mathrm{J})$.
In this context, the following CAV $n$ requires determining a new trajectory by solving an optimization problem that incorporates rear-end safety constraints. 

In such situations, as depicted in Fig. \ref{fig:rear_end_violation_case}, the optimal trajectory becomes a combination of constrained and unconstrained arcs entering and exiting the boundary of the constraint \citep{bryson1975applied}.
Generally, determining the entrance time $t^\mathrm{in}$ and exit time  $t^\mathrm{out}$ can be achieved numerically, but this process involves evaluating all possible combinations, which naturally introduces challenges for real-time implementation.

To address this computational complexity, we present an alternative method.
Suppose we have a candidate trajectory that enters and exits the boundary of the rear-end safety constraints at $t^\mathrm{in}$ and $t^\mathrm{out}$, respectively.
It is crucial to note that this candidate trajectory consists of constrained arcs for $t\in[{t^\mathrm{in}},{t^\mathrm{out}}]$ and unconstrained arcs otherwise, i.e.,
\begin{align}\label{eqn:candidate_u}
    \hat{u}_n(t) = \begin{dcases}
        6\hat{a}_1t+2\hat{b}_1, & \text{if~~} {t_n^0} \leq t < {t^\mathrm{in}},\\
        u_p(t), & \text{if~~} {t^\mathrm{in}} \leq t < {t^\mathrm{out}},\\
        6\hat{a}_2t+2\hat{b}_2, & \text{if~~} {t^\mathrm{out}} \leq t \leq {t_n^f},
        \end{dcases}
\end{align}
\begin{align}\label{eqn:candidate_v}
    \hat{v}_n(t) = \begin{dcases}
        3\hat{a}_1t^2+2\hat{b}_1t+\hat{c}_1, & \text{if~} {t_n^0} \leq t < {t^\mathrm{in}},\\
        v_p(t),& \text{if~~} {t^\mathrm{in}} \leq t < {t^\mathrm{out}},\\
        3\hat{a}_2t^2+2\hat{b}_2t+\hat{c}_2, & \text{if~} {t^\mathrm{out}} \leq t \leq {t_n^f},
        \end{dcases}
\end{align}
\begin{align}\label{eqn:candidate_s}
    \hat{s}_n(t) = \begin{dcases}
        \hat{a}_1t^3+\hat{b}_1t^2+\hat{c}_1t+\hat{d}_1, & \text{if~} {t_n^0} \leq t < {t^\mathrm{in}},\\
        s_p(t)-\delta, & \text{if~~} {t^\mathrm{in}} \leq t < {t^\mathrm{out}},\\
        \hat{a}_2t^3+\hat{b}_2t^2+\hat{c}_2t+\hat{d}_2, & \text{if~} {t^\mathrm{out}} \leq t \leq {t_n^f}.
        \end{dcases}
\end{align}
Here, $u_p(t),v_p(t),s_p(t)$ are the acceleration, speed, and position trajectories of preceding CAV $p$, and $\hat{a}_{(\cdot)},\hat{b}_{(\cdot)},\hat{c}_{(\cdot)},\hat{d}_{(\cdot)}$ are constants.


\begin{proposition} \label{prp:rear1}
    Suppose $\bar{s}_n(t)$ is a cubic polynomial trajectory that connects the points $(t_n^0,0)$ and $(\tau,s_p(\tau)-\delta)$ for $\tau\in(t^\mathrm{in},t^\mathrm{out})$ without violating rear-end safety.
    Then,
    \begin{equation}
        \int_{t_n^0}^{\tau} \bar{u}_n^2(t) dt < \int_{t_n^0}^{\tau} \hat{u}_n^2(t) dt,
    \end{equation}
    where $\bar{u}_n(t)$ is the acceleration trajectory corresponding to $\bar{s}_n(t)$.
\end{proposition}

\begin{proof}
    Recall that the cubic polynomial trajectory in \eqref{eq:optimalTrajectory} satisfies the prescribed conditions \eqref{eqn:lambda}.
    This implies that the optimal trajectory connecting $(t_n^0,0)$ to $(\tau,s_p(\tau)-\delta)$ should inherently take the form of a cubic polynomial unless it violates any of the imposed constraints.
    Hence, it logically follows that if $\bar{s}_n(t)$ does not violate the constraints, then $\int_{t_n^0}^{\tau} \bar{u}_n^2(t) dt$ is the optimal value.
\end{proof}

\begin{proposition} \label{prp:rear2}
    Suppose $\bar{s}_n(t)$ is a cubic polynomial trajectory that connects the points $(\tau,s_p(\tau)-\delta)$ and $(t_n^f,s_n^f)$ for $\tau\in(t^\mathrm{in},t^\mathrm{out})$ without violating rear-end safety.
    Then,
    \begin{equation}
        \int_{\tau}^{t_n^f} \bar{u}_n^2(t) dt < \int_{\tau}^{t_n^f} \hat{u}_n^2(t) dt,
    \end{equation}
    where $\bar{u}_n(t)$ is the acceleration trajectory corresponding to $\bar{s}_n(t)$.
\end{proposition}

\begin{proof}
    The proof follows from a series of arguments analogous to the proof of Proposition \ref{prp:rear1}.
    Thus, the proof is omitted.
\end{proof}

To determine the optimal $t^\mathrm{in}$ and $t^\mathrm{out}$ based on Propositions \ref{prp:rear1} and \ref{prp:rear2}, it is required to identify the largest $t^\mathrm{in}$ that permits a feasible unconstrained arc from $(t_n^0,0)$ to $(t^\mathrm{in},s_p(t^\mathrm{in})-\delta)$ and smallest $t^\mathrm{out}$ allowing a feasible unconstrained arc from  $(t^\mathrm{out},s_p(t^\mathrm{out})-\delta)$ to $(t_n^f,s_n^f)$.
The optimal scenario entails a trajectory where the largest $t^\mathrm{in}$ meets the smallest $t^\mathrm{out}$ at a specific point, i.e., $t^\mathrm{in} = t^\mathrm{out}$.
However, it remains to be established whether such a trajectory exists.


In practical scenarios, a rear-end safety violation usually occurs when the preceding CAV $p$ slows down until $t_p^\mathrm{J}$ to avoid lateral collision and then accelerates to reach its destination by the time $t_p^f$ as shown in Fig. \ref{fig:rear_end_violation_case}.
In such a case, there is a higher chance of unconstrained trajectory from $t_n^0$ to $t^\mathrm{in}$ violating rear-end safety if $t^\mathrm{in} > t_p^\mathrm{J}$ or from $t^\mathrm{out}$ to $t_n^f$ violating if $t^\mathrm{out} < t_p^\mathrm{J}$.

To mitigate this risk, we opt to position the junction point of CAV $n$'s position trajectories at $(t_p^{J},s_p^\mathrm{J}-\delta)$, which is a strategically selected point that potentially offers the most efficient point and feasible unconstrained trajectories.
We verify the feasibility by checking where the constraint boundary intersects with CAV $n$'s trajectory, i.e.,
\begin{equation}
    h(t) := \left(s_p(t)-\delta\right) - s_n(t) \geq 0,~\forall t\in[t_n^0,t_p^\mathrm{J}].
\end{equation}
For $t\leq t_p^\mathrm{J}$, it can be expressed as $h(t)=\left(t-t_p^\mathrm{J}\right)\cdot(at^2+bt+c)$, where $a,b,c\in\mathbb{R}$ are constants, as both arcs $s_p(t)-\delta$ and $s_n(t)$ intersects at the common point $(t_p^{J},s_p^\mathrm{J}-\delta)$.
Consequently, we can simply verify the existence of the roots for $at^2+bt+c=0$ within the range $[t_n^0,t_p^{J})$.
If the trajectory with the junction point $(t_p^{J},s_p^\mathrm{J}-\delta)$ is not applicable, we make adjustments to the flow.

\begin{remark} \label{rmk:collision}
    Generally, an unconstrained energy-optimal trajectory from $(t_p^{J},s_p^\mathrm{J}-\delta)$ to $(t_n^f,s_n^f)$ does not violate rear-end safety because preceding CAV $p$ tends to accelerate faster than CAV $n$ to exit the intersection earlier. In other words, an infeasible solution mostly occurs when an unconstrained energy-optimal trajectory from $(t_n^0,0)$ to $(t_p^{J},s_p^\mathrm{J}-\delta)$ violates the rear-end safety.
\end{remark}

\subsection{Flow Modification}  \label{subsec:flow_modification}

If our planning method produces an infeasible solution, adjusting the flow and re-planing feasible trajectories for the CAVs traveling along their designated routes is necessary.
In this subsection, we present a method that rapidly restores feasibility to the problem.
It is important to note that, in such cases, real-time modification to the flow and adapting trajectories to ensure feasibility is of greater priority than marginal improvement in efficiency.

Let us consider the case where the following CAV $n$ violates rear-end safety for the preceding CAV $p$.
In this context, as highlighted in Remark \ref{rmk:collision}, rear-end safety violation occurs due to the first part of the trajectory, which connects $(t_n^0,0)$ and $(t_p^{J},s_p^\mathrm{J}-\delta)$. 
Therefore, we address the safety violation issue by modifying the first part of the trajectory.

\begin{problem} \label{prb:modification}
    Let $s_n(t):=at^3+bt^2+ct$ be an unconstrained trajectory connecting $(t_n^0,0)$ and $(t_n^c,s_p^\mathrm{J}-\delta)$. Then, we solve the following optimization:
    \begin{equation}
    \begin{aligned}
        \min_{t_n^c} ~~& t_n^c \label{eqn:flow_modification}\\
        \text{s.t. } & s_p(t)-s_n(t)> \delta.
    \end{aligned}    
    \end{equation}
\end{problem}
The coefficients $a,b,c$ are determined by the boundary conditions, i.e., $v_n(t_n^0) = v_n^0$, $v_n(t_n^c) = {v_n^c}^*$, and $s_n(t_n^c) = s_p^\mathrm{J}-\delta$.
For a new trajectory to avoid safety violation, the solution time ${t_n^c}^*$ must be larger than the original time $t_p^\mathrm{J}$. This modification impacts the second part of the trajectory, and thus, we designate the second part as an unconstrained trajectory connecting $({t_n^c}^*,s_p^\mathrm{J}-\delta)$ and $(t_n^f+\tau^\mathrm{new},s_n^f)$, where $\tau^\mathrm{new}=t_p^\mathrm{J}-{t_n^c}^*$.

A similar approach is employed in the case CAV $n$ has a single unconstrained trajectory, but solving the optimization problem for $t_n^f$ instead of $t_n^c$.
Employing this approach allows all CAVs to re-plan and obtain new feasible trajectories.

Meanwhile, the flow should also be updated to enable CAVs departing at a later time to utilize energy-optimal trajectories throughout their trips. We achieve this by updating the free-flow travel time in \eqref{eqn:BPR}, i.e.,
\begin{equation}
\begin{aligned}
t_{ir}^0 &= t_n^c,\\
t_{rk}^0 &= t_n^f+\tau^\mathrm{new},    
\end{aligned}
\end{equation}
where $(i,r)\in\mathcal{E}$ and $(r,k)\in\mathcal{E}$ are the entry road and exit road of the current intersection $r$, respectively.

\begin{figure*}
\begin{tabularx}{\textwidth}{cXX}
\multirow{2}{0.35\textwidth}{\subfloat[Total trips]{\includegraphics[width=\hsize]{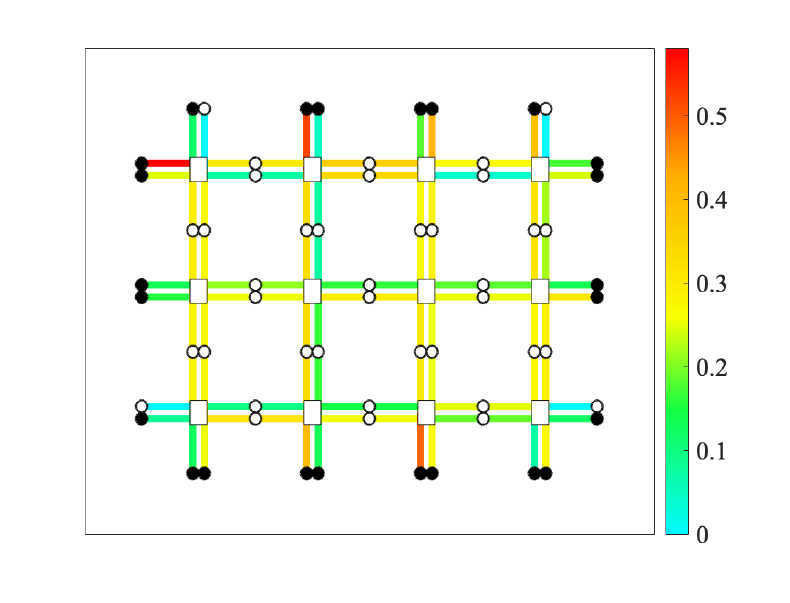}}}
    &\subfloat[Selected trip \#1]{\includegraphics[width=\hsize]{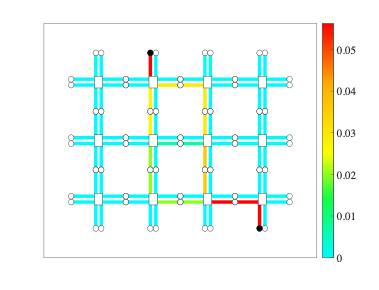}}
    &\subfloat[Selected trip \#2]{\includegraphics[width=\hsize]{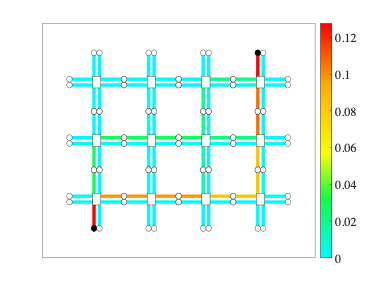}} \\
    &\subfloat[Selected trip \#3]{\includegraphics[width=\hsize]{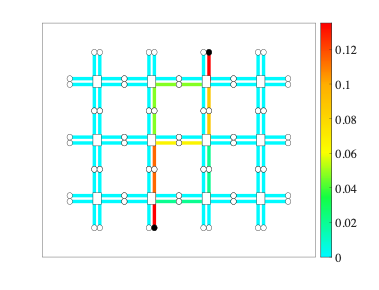}}
    &\subfloat[Selected trip \#4]{\includegraphics[width=\hsize]{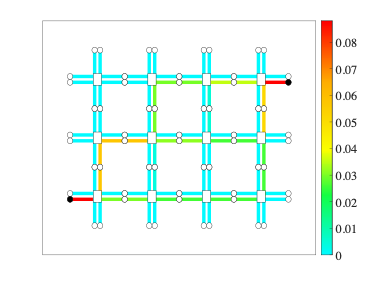}}  
\end{tabularx}   
\caption{Graphical illustration of optimal flow on a grid road network.}
\label{fig:total_flow}
\end{figure*}

\section{Numerical Simulations} \label{sec:simulation}

In this section, we verify the efficacy of our approach through a series of numerical simulations.
First, we concentrated on optimizing the flow of CAVs within a road network comprising $12$ intersections, $62$ depots, and $96$ road segments.
We set each road segment to $200$ meters, resulting in CAVs reaching the next intersection every $400$ meters. We randomly generated $30$ mobility demands to simulate realistic traffic conditions, each characterized by its origin, destination, and demand rate.
As depicted in Fig. \ref{fig:total_flow}, the simulation results illustrate the optimal flow of total and selected mobility demands. The visualization showcases road segments as lines with colors indicating the amount of flow, intersections as white squares, and depots as circles. Black circles represent the origin and destination points of mobility demands. While some roads exhibited congestion due to uneven mobility demands, the overall outcome revealed a well-distributed flow that effectively minimized traffic congestion.

\begin{figure}
    \centering
    \includegraphics[width=0.8\linewidth]{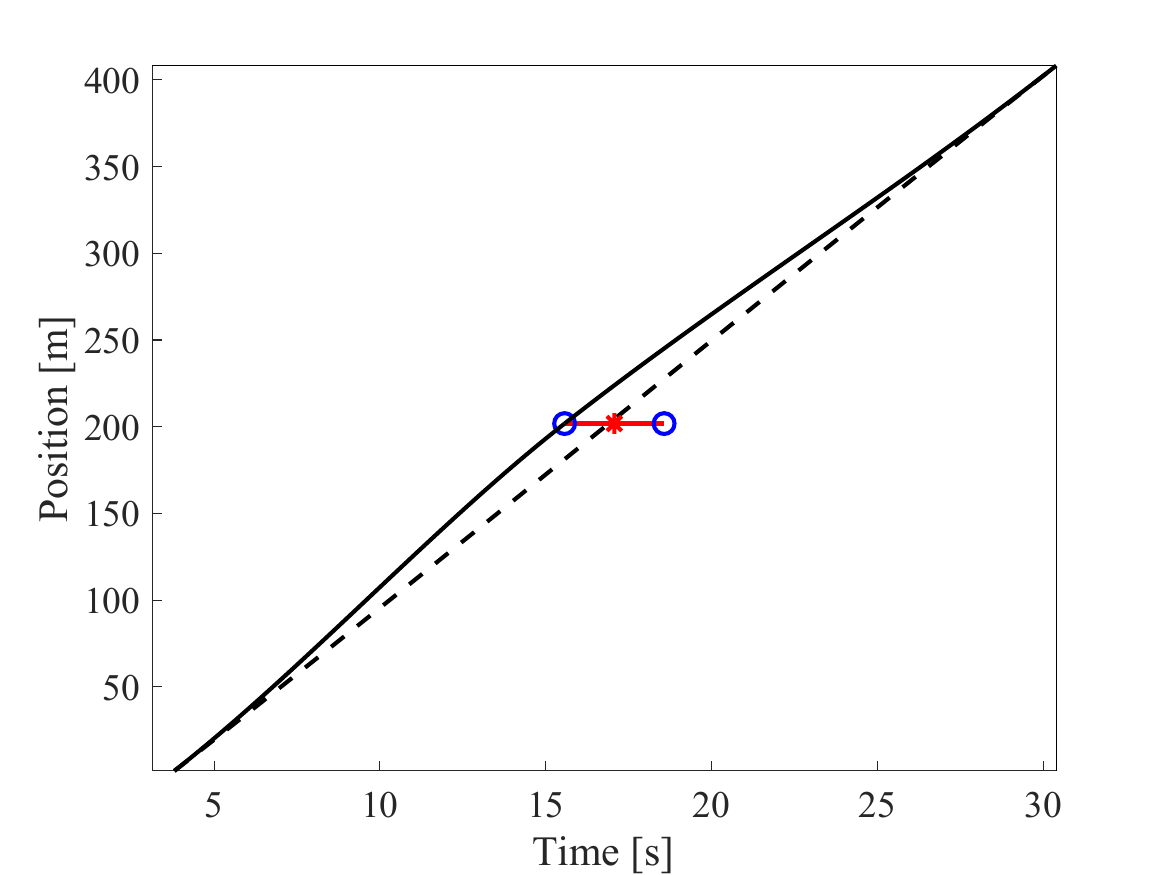}
    \caption{Coordination of a CAV when lateral safety is violated.}
    \label{fig:single_trajectory}
\end{figure}

Next, we conducted simulations of coordination and control of CAVs, focusing on a specific intersection within the network, specifically the bottom left intersection. In this scenario, we introduced $140$ CAVs and generated their entry and exit times based on the method outlined in Section \ref{subsec:route_recovery}.
Figure \ref{fig:single_trajectory} visualizes the trajectory planning of a single CAV. The dashed line represents a single unconstrained energy-optimal trajectory. In contrast, the red point and line depict another CAV navigating a conflict point from a different path and a time headway for lateral safety. The figure shows that, when lateral safety was violated, the CAV used one of the constraint boundaries to piece together two unconstrained energy-optimal trajectories to prevent collisions.

\begin{figure}
    \centering
    \includegraphics[width=\linewidth]{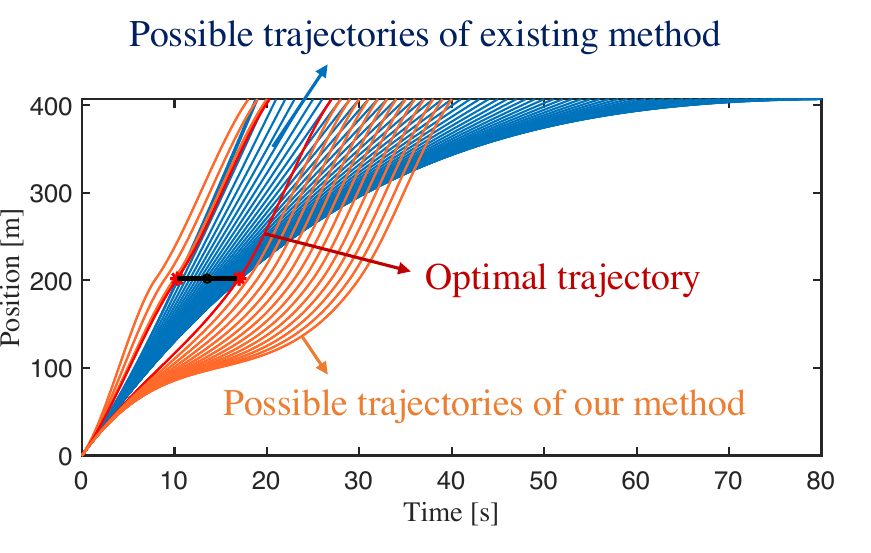}
    \caption{Trajectory comparison between our method and existing method. }
    \label{fig:feasibility}
\end{figure}

Figure \ref{fig:feasibility} presents a comparative analysis of trajectories achievable through our novel approach and the existing method \citep{Malikopoulos2020}.
Blue lines represent potential trajectories from the existing method, while orange lines denote trajectories from our approach. The figure illustrates that our new approach, which strategically selects a junction point connecting two unconstrained trajectories, allows for a more extensive set of feasible trajectories compared to the existing method, which relies on a single unconstrained energy-optimal trajectory.
Notably, our new approach successfully produces an optimal and feasible solution, even when the specific lateral safety constraint (depicted by the black horizontal line in Fig. \ref{fig:feasibility}) renders all possible trajectories from the existing method infeasible.
Furthermore, the trajectories from our approach demonstrate the capability of flow modification. The figure shows that, in cases where flow adjustment is necessary, one can readily choose a feasible trajectory from the possible trajectories (depicted by orange lines) without significantly delaying the arrival time. In contrast, the existing method struggles to provide a feasible trajectory even with substantial delays in the arrival time.

\begin{figure*}
    \centering
    \includegraphics[width=0.85\linewidth]{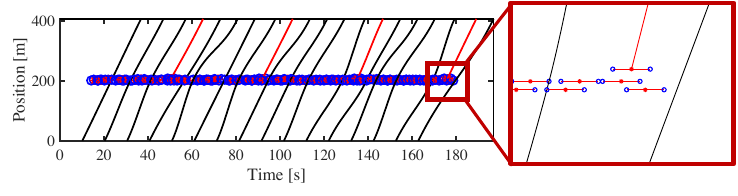}\\
    \includegraphics[width=0.85\linewidth]{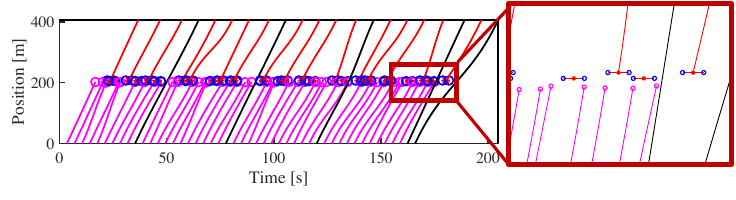}\\
    \includegraphics[width=0.85\linewidth]{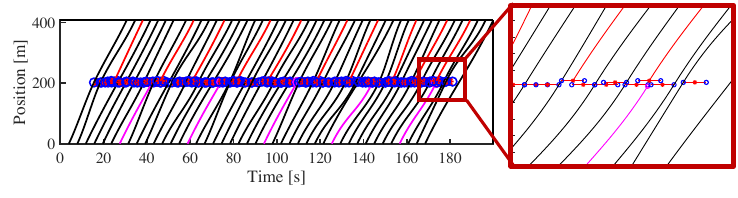}
    \caption{Position trajectories of CAVs using specific paths.}
    \label{fig:coordination}
\end{figure*}

Finally, we conducted simulations at a selected intersection involving 140 CAVs and presented the results in Fig. \ref{fig:coordination}.
Specifically, we focused on the left-bottom intersection in the grid network (depicted in Fig. \ref{fig:total_flow}), which includes two origins and one destination of trips.
Each graph in Fig \ref{fig:coordination} represents a distinct entry and exit scenario, showing the trajectories of CAVs passing through their designated path.
The red and purple trajectories indicate CAVs merging with and diverging from the specified path, causing trajectories to appear or disappear at conflict points.
The figure vividly illustrates that CAVs efficiently planned their trajectories to cross the intersection one after another by adhering to the boundaries of safety constraints.
Consequently, the majority of CAVs efficiently filled the gap, enabling the introduction of a high density of CAVs in a single intersection. Those CAVs that do not follow the boundary of safety constraints selected the unconstrained energy-optimal trajectory.
In conclusion, the results confirm that our framework selected appropriate exit times based on the flow, minimizing cascading delays and ensuring smooth traffic flow for widely and densely distributed CAVs.

\section{Concluding Remarks} \label{sec:conclusion}

This paper proposes a hierarchical decision-making framework of emerging mobility systems for optimal operational strategies at various levels. Our approach addressed challenges related to combining on-demand flow optimization, route recovery problems, and vehicle-level trajectory planning at intersections.

At a macroscopic level, we addressed the on-demand flow optimization problem. We provided a heuristic method to efficiently extract routes from the optimal flow while assigning suitable departure times for CAVs.
At a microscopic level, we focused on solving coordination problems at intersections, ensuring interaction among CAVs while preserving the overall optimal flow.
We established the optimality conditions for vehicle-level trajectory planning and proposed a rapid computation method for finding optimal trajectories.
Additionally, we provided an adaptive method to modify flow and trajectories, ensuring continuous feasibility for CAV solutions.

While our current study primarily delved into grid networks with intersections, the real-world landscape introduces various scenarios, such as merging roadways, highways, and signalized intersections.
Therefore, as future work, verifying our method in realistic traffic scenarios would be worthwhile.
Furthermore, extending our framework to accommodate mixed traffic environments, where CAVs dynamically plan trajectories alongside human-driven vehicles, presents an intriguing avenue for future research in intelligent transportation systems.



\bibliographystyle{abbrvnat}
\bibliography{Bang, IDS}

\end{document}